
  \input amstex
  \documentstyle{amsppt}
  \par
  \null
  \magnification=\magstep1
  \hsize=16truecm
  \vsize=24truecm
  \voffset=0
  \baselineskip
  \par\noindent
  \tolerance=10000
  \def\cT{\Cal T}

  \def\cM{\Cal M}
  \def\cX{\Cal X}
  \def\cK{\Cal K}

  \def\bY{\bar Y}
  \def\bxi{\bar \xi}
  \def\bpsi{\bar \psi}
  \def\bX{\bar X}
  \def\bt{\bar \tau}
  \def\bD{\bar D}
  \def\S{\bold S}
  \def\Z{\bold Z}
  \def\T{\bold T}
  \hyphenation{ma-ni-fold La-gran-gi-ans}
  \par
  \line{\hfil {\it Talk given at the Fourth Hungarian Relativity Workshop,}}
  \par
  \line{\hfil {\it org. R.P. Kerr and Z. Perj\'es, G\'ardony, 12-17 July 1992}}
  \par
  \line{\hfil Preprint NSI-8-93; gr-qc/9303003 }
  \bigskip
  \topmatter
  \title  STRING-LIKE STRUCTURES IN COMPLEX\\
  KERR GEOMETRY
  \endtitle
  \author
  A. Ya. Burinskii
  \endauthor
  \affil
  Nuclear Safety Institute of the Russian Academy of Sciences,
  B.Tulskaya 52  Moscow 113191 Russia, e-mail: grg\@ibrae.msk.su
  \endaffil
  \abstract
  The  Kerr geometry is represented as being created by a source moving
  along an analytical complex  world-line. The equivalence of this complex
  world-line and an Euclidean version of complex strings (hyperbolic
  strings) is discussed.
  It is shown that the complex Kerr source satisfies the corresponding
  string equations. The boundary conditions of the complex Euclidean
  strings require an orbifold-like structure of the world-sheet. The
  related orbifold-like structure of the Kerr geometry is discussed.
  \endabstract
  \endtopmatter
  \document
  \bigskip
  {\bf $1$.  Introduction}
  \medskip
  \par
  Much attention has been paid recently to the relations of the two-dimensional
  black holes and strings [1]. In this paper we shall consider the
  four-dimensional  Kerr geometry and  derive some
  string-like structures  for its complex source.
  \par
  The Kerr
  solution [2] is well known as the  field  of a rotating black hole. However,
  for the case of  a  large  angular momentum  $\mid a\mid  = L/m \geq  m$
  the its horizons are all absent and a  naked ring-like  singularity appears,
  along which space branches into two sheets.
  \par
  This naked singularity has many unpleasant manifestations, and
  must be hidden  inside a rotating disk-like source [3,4]. The
  solution with $\mid a\mid  \gg  m$  displays some remarkable
  features suggesting certain relationships with the spinning elementary
  particles [3-6].
  The real structure of the Kerr metric source has itself some
  quite exotic properties [4] and contains a string-like closed vortex line
  resembling the superconducting strings of Nielsen-Olesen and Witten [7].
  \par
  The aim of this paper is to show some connections between the
  complex structure of the Kerr geometry and an Euclidean version of
  string theory. We find, that from a complex point of view the source of
  the Kerr geometry represents
  a complex Euclidean string propagating in  complex Minkowski space.
  \par
  We start from  the complex structure of the Kerr
  geometry and describe the complex world-line representation [8]. The Kerr
  geometry can then be represented as a retarded time solution  generated by
  a source moving along  a complex  world  line.
  We use the Kerr-Schild formalism [2], which is close
  to the geometrical twistor description.
  The necessary conditions for the nonstationary generalization of the
  complex world-line representation are obtained.
  \par
  The complex world line is really a two-dimensional surface,
  equivalent a two-dimensional field theory  [9], and may be considered
  as an intermediate object between
  a particle and a Euclidean version of a string [10-12].
  The peculiarities
  of these Euclidean strings are analyzed, and it is shown that the
  boundary conditions of the complex Euclidean string require an
  orbifold-like structure of the world-sheet.
  The corresponding orbifold structure of the complex Kerr geometry is
  discussed.
  It links the retarded and advanced folds of the complex light cone, which
  plays a very important role in the structure of the Kerr geometry.
  Such an orbifold-like structure was recently suggested for two-dimensional
  black holes by Witten [1].
  \par
  In the last section we shall briefly discuss  some generalizations.
  \bigskip
  {\bf $2$. The complex world-line representation of the Kerr geometry}
  \medskip
  \par
  We  use the Kerr-Schild form of the Kerr-Newman metric [2]
  $$ g_{ik} = \eta_{ik} + h k_{i}k_{k}; \eqno(2.1) $$
  where $\eta _{ik}$=diag(-1,1,1,1) is the
  auxiliary Minkowski metric in Cartesian coordinates $(x,y,z,t)$, and
  $ h=(2mr-e^{2})/(r^{2}+ a^{2}\cos ^{2}\theta ). $
  The vector null field $k_{i}(x)$ is a geodesic and shear-
  free principal null congruence and may be represented
  in  the form  $$ K(x) = k_{i}dx^{i}=
  P^{-1}( du +\bY d\xi + Y d \bxi - Y \bY dv ), \eqno(2.2) $$
  where the coordinates used are null
  $ u=(z+t)/ \sqrt{2};\quad
  v=(z-t)/\sqrt{2}; \quad \xi =(x+iy)/\sqrt{2};\quad \bxi=(x-iy)/\sqrt{2},$
  and  $Y$ is a complex projective spinor coordinate,
  $ Y = \psi ^{1}/ \psi ^{0}. $
  The complex world line representation was first introduced by Lind
  and Newman  [8]. It shows  that
  the Kerr-Newman solution may be considered as a retarded time solution
  generated by  a "complex point" source [8,6]. The source propagates
  in  complex Minkowski space $CM^4$
  along a complex "world line"
  $X^{i}_{o}(\tau),\quad (i=0,1,2,3),$
  parametrized by a complex time parameter
  $\tau = t+i\sigma  = X^{o}_{o}(\tau) .$
  An important role in this retarded time construction
  is played by the complex light cones $\cK $,
  whose apexes lie  on  the  complex "world line" $X_{o}(\tau)$.
  \bigskip
  {\bf 2.1 Complex light cone}
  \medskip
  \par
  The complex light cone $\cK$, written in spinor form  $$ {\cK }= \{x: x =
  X^{i}_{o}(\tau) + \psi ^{A}_{L} \sigma ^{i}_{A \dot { A}} \tilde{\psi
  }^{\dot{A}}_{R} \} \eqno(2.3)$$ \noindent may be split  into
  two families of null planes: "left" $( \psi _{L}$ =const; $\tilde{\psi
  }_{R}$ -var.) and "right"$( \tilde{\psi }_{R}$ =const; $\psi _{L}$ -var.).
  These are the only two-dimensional planes which are wholly contained in the
  complex null cone.
  The rays of the principal null congruence $K$ of the Kerr geometry  are
  the tracks of these complex null planes (right or left) on the real slice of
  Minkowski space.
  The light cone equation in the Kerr-Schild metric coincides with the
  corresponding equation in Minkowski space because the null directions
  $k^i $ are null in both metrics, $g_{ij}$ and $\eta _{ij} $.
  \par
  This splitting of the light cone on the null planes has a close
  connection with the twistors. By introducing the projective
  parameter $Y = \psi ^{1}/ \psi ^{0}$ the complex light cone equation
  $$ (\xi  -\xi _{o}) (\bxi -\bxi _{o}) = - (u - u_{o}) ( v - v_{o}) $$
  splits into two linear equations \footnote{It is a generalization of
  the Veblen and Ruse construction [13] which has been used for the
  geometrical representation of spinors.} $$\eqalign{ \xi  - \xi _{o}
  &= Y (v - v_{o} ), \cr -Y (\bxi - \bxi _{o} ) &= (u - u_{o} ),}
  \eqno(2.4) $$ \noindent describing the "left" complex null planes (
  the  null  rays  in  the real space).  Another splitting $$\eqalign{
  - \tilde {Y} (\xi  - \xi _{o}) &=  (u - u_{o} ), \cr (\bxi - \bxi
  _{o} ) &= \tilde{Y} (v - v_{o} ),} \eqno(2.5) $$ \noindent gives the
  "right" complex null planes.  The three parameters $$ Y,\qquad
  \lambda _{1} = u + Y \bxi , \qquad \lambda _{2} = \xi - Y v
  \eqno(2.6) $$ \noindent are indeed projective twistor
  coordinates.\footnote{In the standard twistor form $ (1, Y, \lambda
  _{1} , \lambda _{2}) = (Z^{o}, Z^{1}, Z^{2}, Z^{3})/Z^{o} , $  where
  $  Z = (\psi ^{B} , \pi _{\dot A}) ; \qquad \pi _{\dot A} = X _{\dot
  A B} \psi ^{B}. $} The equations for the "left" null planes of the
  complex light  cone with apex at the point $X_{o}$ take the form  $$
  \lambda _{1} = \lambda ^{o}_{1} = u_{o} + Y\bxi _{o} , ,\qquad \lambda _{2}
  = \lambda ^{o}_{2} = \xi _{o} - Y v_{o}.   \eqno(2.7) $$
  \noindent The "left"
  null planes of the complex light cones form a  complex congruence
  or regulus which generates all the rays of  the  principal  null
  congruence $K$ in real space. The ray with polar direction $\theta,\phi$
  is the real track of the "left" plane corresponding to $Y = \exp{i \phi}
  \tan (\theta /2) $ and belonging to the cone which is placed at the point
  $X_o$ corresponding to  $\sigma = a \cos (\theta).$ The parameter $\sigma
  = Im \tau$ has a meaning only in the range $- a \leq \sigma \leq a$ where
  the cones have  real slices.
  \par
  Thus, the complex world line
  $X_{o}(t,\sigma )$ represents a restricted two-dimensional surface
  or strip, in  complex Minkowski space, and is really a world-sheet.
  It may  be considered  as a complex open string with a Euclidean
  parametrization $\tau = t+i\sigma ,\bt =t-i\sigma $,  and with end
  points $X_{o}(t,\pm a)$.
  \bigskip
  {\bf 2.2 The complex world lines and the open Euclidean strings}
  \medskip
  \par
  The complex world line, or world-sheet, is an intermediate object
  between particle and string  [10,11]. \footnote{The
  corresponding complex Euclidean
  strings have  also been considered in recent papers of
  Oogury and Vafa [12], where they  were called "hyperbolic".}
  The complex source of the Kerr-Newman solution corresponds
  to a straight analytical world-line in the complex Minkowski space $CM^4$
  $$ X^{i}_{o}(\tau) = X^{i}_{o}( 0 ) + U^{i}_{o} \tau .
  \eqno(2.8)$$
  In the Euclidean version, the general string solution has to be a sum of
  the left and right modes, taking the form
  $$ X^{i}(t,\sigma ) = X^{i}_{L}(\tau ) + X^{i}_{R}(\bt ).
  \eqno(2.9)$$
  It seems  most natural to use the complex conjugate world-line as the
  right mode,  leading to the real string solution
  $ X^{i}(t,\sigma ) $
  However, the string solutions have also to satisfy  the constraints
  $$ (\partial
  _{\tau}X)^{2}= 0;\qquad (\partial ^{\_}_{\tau}X)^{2}= 0,
  \eqno(2.10)$$
  where  $$ \tau = t + i\sigma;\quad \bar{\tau} = t - i\sigma  ;\qquad
  \partial_{\tau} = (\partial _{t} - i\partial _{\sigma })/2 ;\qquad
  \partial^{\_}_{\tau} = (\partial _{t} + i\partial _{\sigma })/2,
  \eqno(2.11)$$
  and the boundary conditions
  for $\sigma  = \pm  a$ are  $$ Re{[
  {{\partial  L_{e}}\over { \partial (\partial_{\tau} X^{i}) } }]}_{\mid
  \sigma =\pm a} = 0. \eqno(2.12)$$
  This equations puts certain strong restrictions on the solutions, with the
  result that the real solutions for the open Euclidean strings are absent.
  Nevertheless, one can find a consistent formulation of the complex
  closed Euclidean strings, in which the complex analytical straight
  worldlines  form  closed strings satisfying the  corresponding full
  system of string equations.
  \bigskip
  {\bf $3$.  Closed complex Euclidean strings with an orbifold-like
  world-sheet}
  \medskip
  \par
  One can consider Euclidean strings in complex Minkowski space $CM^4$
  as  a complex objects with  a Hermitian Lagrangian,
  $$ L = - \eta _{i\bar {j}} (\partial _{\tau } X^{i}
  \partial _{\bt } \bX ^{\bar {j}} + \partial _{\bt } X^{i} \partial _{\tau}
  \bX ^{\bar {j }}), \eqno(3.1)$$
  The general solution is given as a sum of the right and left modes,  $$
  X^{i}(t,\sigma ) = X^{i}_{L}(\tau ) + X^{i}_{R}(\bt ),
  \eqno(3.2)$$ \noindent
  which are not necessarily complex conjugates of each other.
  The constraints take the form
  $$ \eta _{i\bar j} \partial _{\tau} X^{i}_{L} \partial _{\tau}{\bX
  }^{\bar j}_{R} = 0; \qquad \eta _{i\bar j} \partial _{\bt }
  X^{i}_{L} \partial _{\bt }{\bX }^{\bar j}_{R} = 0.   \eqno(3.3)$$
  An oscillator  expansion of the Euclidean string is
  $$\eqalign{X_{L} &= X_{L}(0) +
  P\tau + \omega ^{-1}\sum (1/n) \alpha _{n} \exp \{ -i n \omega  \tau
  \} , \cr
  X_{R} &= X_{R}(0) + P\bt + \omega ^{-1}\sum
  (1/n) \tilde {\alpha }_{n} \exp \{ -i n \omega \bt \} ,} \eqno(3.4)$$
  \noindent where a
  periodic time dependence is proposed with a period $T = 2\pi /\omega $.
  The expansion contains the hyperbolic basis function, which are not
  orthogonal over the string length.
  For the parameter $\sigma$ it is convenient to use an interval
  $[-a, a]$.  \noindent The complex Euclidean strings can  not  be
  open, since the boundary conditions for open strings,  $\qquad
  (\partial _{\sigma } X)^{2}_{\mid \sigma =\pm a} =0 \qquad$ lead only
  to a trivial solution.
  \par
  Attempts to introduce boundary conditions  for  closed complex Euclidean
  strings meet obstacles too, because it  is impossible to introduce
  the same boundary conditions for the real and imaginary part of a string.
  However, the problem may be resolved  when the world sheet admits an
  orbifold-like structure [11,14,15]. The general idea is very simple:
  it is necessary to modify the world-sheet to include the twisting boundary
  conditions for the imaginary part of the complex string.
  A more precise formulation is given by the following.
  \par
  To construct an orbifold
  it is necessary to double the interval $\Sigma =[-a, a]$   for the
  parameter $\sigma = Im \tau$ .
  Two copies of the interval,  $\sigma _{+} \in
  \Sigma _{+} = [-a, a]$ and  $\sigma _{-} \in
  \Sigma _{-} = [-a, a]$,
  must  be joined  to  form  an oriented
  circle, ${\S}^{1}= \Sigma _{+} \cup \Sigma _{-}$ which may  be
  parametrized  by  a  periodic coordinate $\theta  \sim \theta + 2\pi $,
  and represents  a covering space for the
  original interval $\Sigma $ .
  For an explicit correspondence one can use the projection $\S ^1$
  on $\Sigma$, $$\sigma = a \cos \theta . \eqno(3.5)$$
  The original string is then parametrized by $0 \leq \theta \leq \pi$,
  and $\pi\leq \theta \leq 2\pi$ covers the string a second time in the
  opposite direction.
  \par
  A group $ G ={\Z}_2; \quad  g\in G,$ acts on the circle
  as $g \theta  = \theta  + \pi ; \qquad
  g^{2}= 1$, and induces the transformation $$ g \sigma _{+} = - \sigma _{-}
  ;\qquad g \sigma _{-}= - \sigma _{+};\qquad g \tau _{+}= \bar {\tau}_{-}
  ;\qquad g \tau _{-}= \bar {\tau} _{+}.   \eqno(3.6) $$
  The string modes $X_L (\tau _L)$ and $X_R (\bt _R)$ can
  be extended on the circle for $\pi \leq \theta \leq 2\pi $ to form a closed
  string  by means the well-known kind of extrapolation [15]
  $$ X_L (g\tau _+) = X_R(\bt _+); \qquad
  X_R (g\bt _+) = X_L(\tau _+)  \eqno(3.7) $$
  The group G is a symmetry of the theory and therefore acts on the string
  variables,
  $$\tilde {g}X_L(\tau _\pm) =X_L (g\tau _\pm) = X_R (\bt _\pm),
  \qquad \tilde{g}X_R(\bt _\pm) =X_R (g\bt _\pm) = X_L (\tau _\pm).$$
  One can see that the group acts by interchanging the left and right modes.
  The  time coordinate $t$ may be transformed by projection on  the circle
  ${\S}^{1}$ so that the world-sheet
  forms the torus ${ \T}^{2} = {\S}^{1} \times  {\S}^{1}$ .
  The action of the time reversing  element $h, $ of ${\Z}_{2} $
  is $h t = - t, \quad h^2 =1$. The elements $g$ and $h$  commute with
  each other and act together as
  $gh \tau_{L} = - \tau_{R}; gh \tau_{R}= - \tau_{L};\qquad
  h \tau_{L} = -g \tau_{R}; h \tau_{R}= -g \tau_{L}$.
  The quotient space
  ${\T}/{\Z}_{2}$  is the orbifold.
  The action $g$ on the orbifold is defined by the projection (3.5)
  $$ g\sigma= - \sigma, \quad g\tau = \bt,  \quad \tilde g X(\tau)
  = X(g\tau) = X (\bt).  \eqno(3.8)$$
  The actions of  $gh$ and $h$ are
  $$ gh \tau = - \tau,\quad gh \bar{\tau} = - \bar{\tau}\quad,\quad
  h \tau = - g\tau,\quad h \bar{\tau} = - g\bar{\tau}. \eqno(3.9)$$
  \par
  The Hilbert space of the string solutions H on the orbifold have to be
  decomposed  on two subspaces: the $H_{+}$  space of the even functions,
  $$ X_{ev} (t,\sigma ) =
  ( X + \tilde g X)/2 , \qquad \tilde g X_{ev} =
  X_{ev}, \eqno(3.10)$$
  and the $H_{-}$  space of the odd functions,
  $$ X_{odd} (t,\sigma ) =
  (X - \tilde g X)/2;\qquad\tilde g X_{odd} = - X_{odd}.\eqno(3.11)$$
  As a result the even solutions are closed on the orbifold and have the
  form
  $$ X_{ev}= ( X_L(\tau) + X_R(\bt ) + X_L(\bt ) + X_R(\tau ))/2,
  \eqno(3.12)$$
  whereas the odd solutions,  $$ X_{odd}=
  ( X_L(\tau ) + X_R(\bt ) - X_L(\bt ) - X_R(\tau ))/2,  \eqno(3.13) $$
  form the closed strings on the orbifold with twisted boundary conditions.
  A general complex string solution on the orbifold may be represented as
  the sum of the even and odd parts
  $$ X(t,\sigma ) = X_{ev} + X_{odd} =
  (X + \tilde g X)/2 + (X
  - \tilde g X)/2 .  \eqno(3.14)$$
  The $gh$  transformations  do not mixes  the  analytic and antianalytic
  parts  of  the  functions  on  the  orbifold,  but the functions may be
  decomposed into the even and odd parts with respect  of the $gh$
  transformations $ X(\tau) = X_{gh+}(\tau) + X_{gh-}(\tau ) .$
  In the terms  of  the $gh$  decomposition  the  general  even  and
  odd solutions for a closed string are
  $$X_{ev}= 2Re X_{gh+} + 2i Im X_{gh-} ; \quad
  X_{od}= 2Re X_{gh-} + 2i Im X_{gh+} .\eqno(3.15)$$
  \par
  The complex constraints (3.3) then take the form
  $$ g_{i j}[\partial _{\tau}Re X^{i} \partial _{\tau}{ReX
  }^{ j}+ \partial _{\tau}Jm X^{i} \partial _{\tau}{JmX
  }^{ j}] = 0  ,  \eqno(3.16$$
  which is a generalization of the real constraints, taking into account that
  the imaginary components are the independent degrees of freedom.
  \par
  These constraints admitt one interesting class of  solutions
  with $X_R = 0$,
  $$X^{i}(t,\sigma ) = X^{i}_{L}(\tau), \eqno(3.17) $$
  which can be seen from the complex form of constraints (3.3).
  These solutions contain only left modes like the Schild null
  strings or the bosonic sector of the heterotic string, and correspond
  to analytical world-sheets.
  In spite of the absence of a right mode the solutions are nontrivial and
  contain both even and odd parts. It is interesting that the straight
  analytical world-line corresponding to the stationary Kerr
  solution satisfies
  the constraints as well as the arbitrary analytical complex
  world-lines. \footnote{In the case
  of real strings the stationary solutions do not satisfy the constraints.}
  Thus, the world-line corresponding to the complex
  source of the Kerr solution satisfies the string
  equations on the orbifold.
  \par
  \bigskip
  {\bf $4$.  Orbifold-like structure of the complex Kerr geometry}
  \medskip
  \par
  To obtain a real cut of the complex Kerr-Schild space for every $\tau$
  we shall use a null vector
  $$
  {\cM }^{i} = \psi  \sigma ^{i}\tilde {\psi }
  \eqno(4.1)$$
  \noindent of the complex light cone, linking  points  of  the  complex
  world line $X_{o}(\tau)$ with  points of the real slice. If a cone has  a
  real slice, than for every normalized  spinor $\psi$
  (with  $\psi \bar {\psi } =1 $)  there exists a spinor
  $$
  \tilde {\psi } = \bar {\sigma }^{o}\sigma _{a}( X^{a}_{o}- \bar {X}^{a}_{o})
  \bar {\psi },\quad (a=1,2,3), \eqno(4.2)$$
  \noindent such that $X_{o}+ {\cM }$ is a point on the real slice, i.e.
  $$ X_{o}+ {\cM } = \bX _{o}+ \bar {\cM}   .   \eqno(4.3)$$
  The explicit form of ${\cM }$ is the following
  $${\cM }^{i}= (X^{i}_{o} -\bX ^{i}_{o})/2 + i \epsilon^{iojk} k_{k}
  (X_{oj} -\bX _{oj} )/2 , \eqno(4.4)$$
  \noindent where $k_{i}= \psi  \sigma _{i}\bpsi $ is  the  real  vector  of
  the  principal  null congruence.
  One can also add to  $\cM $ a real vector directed along the null ray
  $ r  k^i$. The null vectors $ k^{i} $ and $ \cM ^{i}$ belong to the "left"
  null plane.
  \par
  One can find even and odd structures in the Kerr geometry.
  The real "center" of the Kerr geometry, $ X_{re} (t,\sigma )
  = ( X_{oL}(\tau )+ \bar X_{oL}(\bt ))/2$, is an even object.
  The imaginary part of the world-line $ X_{im} (t,\sigma)
  = ( X_{oL}(\tau ) - \bar X_{oL}(\bt ))/2$
  has odd parity with respect to complex time, but it also contains
  an even constant $ X_{oL}(0)$.
  The imaginary degrees of freedom in the Kerr geometry may be
  included in a compactification scheme based on
  the constrained $CP^3$ sigma-model [16].
  The light cones, which are  adjoined  to every  point of
  the complex world line, can be considered as an "internal"  space  of the
  Kerr string in analogy with the compactification in string models.  \par
  One can introduce relative coordinates for the real points of the
  Kerr-Schild space, $$
  Z^{i} = X^{i} - X^{i}_{o} ,\quad  i=0,1,2,3, \eqno(4.5)$$
  which include the imaginary part of the complex world-line
  $Im Z^{i} = - Im X^{i}_{o}$
  and satisfy the light cone constraints,  $$ Z_{i}Z^{i} = 0 ,
  \eqno(4.6)$$
  In the constrained sigma-model
  the $Z^{i}$ are considered as homogeneous coordinates for
  the complex projective space ${\bold CP}^{3}$, $$\xi ^{\alpha } = Z^{\alpha
  } / Z^{o} ,\qquad \alpha  = 1,2,3.\qquad \eqno(4.7)$$
  The space may be endowed
  with a K\"ahler metric in the standard manner,
  $$ g_{\alpha \bar \beta
  } = \partial _{\alpha } \partial _{\bar \beta } {\cM }^{fs},
  \eqno(4.8)$$ \noindent
  It is a metric on the compact space of the  complex null rays
  defined by the Fubini-Study potetial,
  $$ {\cM }^{fs} = \ln (1 + \xi \bxi ).  \eqno(4.9)$$
  For every real point of the Kerr-Schild space $X^i$ one can find
  both the "left" retarded time $\tau_L$ and a point of the complex world-line
  $X_o (\tau _L)$ defined by the intersection of the world-line
  with the "left" null plane (2.4).
  This plane is spanned by  the null
  vectors $\cM ^i $ and $k^i$. The "right" plane (2.5) contains the vector
  $\cM ^i$, but it also defines the congruence $l ^i= \tilde {\bpsi }
  \sigma ^i\tilde {\psi } $ through the equation
  $\tilde {Y}= \tilde {\psi^1} /\tilde {\psi ^0}$. It should be mentioned
  that the "right" and "left" planes change places under the action of the
  gh - transformation $ (v -v_o) \longleftrightarrow (u-u_o)$
  and a redefinition $\tilde {Y} = -1/Y$.
  On the other hand the gh-transformations change the retarded and advanced
  folds of the complex light cone,
  as  can be seen from the alternative splitting of the complex light cone,
  $$ t-\tau = \pm \tilde { r} ;\qquad gh \tilde { r} = - \tilde {r},
  \eqno(4.10)$$
  where $ (\tilde {r})^2= (x-x_o)^2+(y-y_o)^2 +(z-z_o)^2 $
  When  $t$ is real, the transformation g acts on $\tilde {r} $
  as complex conjugation.
  These transformations induce the orbifold structure on the $CP^3$ by
  defining an equivalence between positive and negative folds relating $Z^0$
  to $-Z^0= gh Z^0$, and by introducing two new folds of the orbifold
  corresponding to $\pm \bar {Z^0} = g(\pm Z^0)$.
  This orbifold may be described in $CP^3$ by the quartic equation
  $$ [Z_1^2 +Z_2^2+Z_3^2 -{g Z_0^2}] (Z_1^2 +Z_2^2+Z_3^2 -Z_0^2) =0 .
  \eqno(4.11)$$
  This equation corresponds to a peculiar case of the $K3$ surface, a known
  example of the Calaby-Yau manifolds [14,18].
  Notice, that the complex light
  cone plays an important role in  the structure of the Calabi-Yau spaces.
  It is a "conifold" [17], or a singular limit of  many Calabi-Yau
  spaces. The procedure of "small resolution" of the
  singularities is produced by introducing the extra $CP^{1}$
  space $Y = \psi ^{1}/ \psi ^{0}$ and splitting the constraints [17].
  One can note, that this is an  exact copy  of  the above
  (see eq.(2.4), (2.5)) splitting of the complex light cone on the left
  (or right) twistor null
  planes. It not only removes  the singularity but  also  annihilates  the
  multiple connectedness caused by the group of discrete symmetry.
  \bigskip
  {\bf $ 5$.  Generalizations}
  \medskip
  \par \noindent
  We will discuss two generalizations of the previous sections:
  \bigskip
  {\bf 5.1 A nonstationary generalization of the complex world-line
  representation}
  \medskip
  \par
  The complex Kerr source corresponds to the straight analytical word-line,
  but from the point of view of string excitations the oscillating
  world  lines are  very interesting.
  The problem of the existence of corresponding
  solutions is open. One of the problems involved in obtaining the
  nonstationary generalizations of the Kerr-Schild solutions is the
  question of the existence of the corresponding principal null congruencies
  (which  have to be both geodesic and shear-free).
  \par
  In this section we shall briefly discuss the nonstationary generalization
  of congruencies, following [20].
  According to Kerr's theorem [2,19] the geodesic and shear-free congruencies
  are defined by eq.(2.2), where the
  function $Y(X)$ is a solution of the equation $$ F(Y, \lambda
  _{1},\lambda _{2}) = 0, \eqno(5.1)$$ \noindent
  $F$ being  an analytical  function.  The singularities  of  the
  solutions occur as caustics
  in $Y$, satisfying the equations $ F = 0; \qquad \partial _{Y} F = 0. $ In
  the case when the singularities  are  confined to   a   bounded region
  [21,6] the $F$ is, at most, quadratic in $Y$  and the equation $F(Y)=0$
  may  be solved explicitly, giving an explicit form for the
  principal null congruence.  The function F may be
  represented in a form
  $$ F= (\partial _{\tau} \lambda ^{o}_{1}) (
  \lambda _{2}- \lambda ^{o}_{2}) -(\partial _{\tau} \lambda ^{o}_{2}) (
  \lambda _{1} - \lambda ^{o}_{1} ),  \eqno(5.2)$$ which is
  convenient for the nonstationary generalizations because  the parameters
  $\lambda ^{o}_{A}$   and  $\partial _{\tau} \lambda ^{o}_{A}$
  are  the functions of the complex world line parameters
  $X_{o}(\tau)$ and therefore depend on $\tau$ .  \par
  Following the general scheme described in [2], one can show that
  the shear-free and geodesic conditions for the congruence
  $$ ( \partial _{\bxi } - Y
  \partial _{u}) Y = 0;\qquad (\partial _{v} + Y \partial _{\xi } +\bY
  \partial _{\bxi } - Y\bY\partial _{u}) Y = 0.  \eqno(5.3)$$
  lead  to a differential equation, $$ Z^{-1}(\bY - \phi )dY = \phi
  (d\xi  - Y dv) + (du + Y d \bxi) , \eqno(5.4)$$ \noindent where $Z
  = ( \partial _{\xi } - \bY \partial _{u}) Y$    is the  complex  expansion
  of  the congruence, and $\phi $ is an arbitrary solution of the equation
  $$ ( \partial ^{\_}_{\xi } - Y \partial _{u}) \phi  = 0.
  \eqno(5.5)$$
  By using the form (5.2) one can obtain the following equations which are
  necessary for the congruence to be both geodesic and shear-free,
  $$ ( \partial ^{\_}_{\xi } - Y
  \partial _{u}) \tau = 0;  \eqno(5.6)$$
  \par
  $$ \lambda _{1} = \lambda^{o}_{1} (\tau) ;
  \qquad \lambda _{2} = \lambda ^{o}_{2}
  (\tau).  \eqno(5.7)$$
  These equations may be considered as
  constraints on the  retarded time parameter $\tau$.  Equations
  (5.7) are equivalent to (2.4) and imply that the
  points $X$ and $X_{o} (\tau)$ belong to the same "left" null plane of the
  complex light cone.  Consequently, the retarded time $ \tau = \tau _L(X)$
  have to be defined by the point of intersection of the "left" null plane
  with the world line $X_{o}$, or by the "left"
  projection of the X on the world line along the corresponding "left" null
  planes.  Thus, in the nonstationary case the complex world line
  representation acts via the "left" retarded time $\tau _L$, which must
  be used in the parameters of equation (5.2).  \par Equation
  (5.6) gives an extra necessary condition on the world-line
  $X_{o}(\tau)$. It has to be an analytical function of $\tau$.
  Remarkably, this condition coincides with the necessary condition on the
  world-line to be a solution of the complex Euclidean string equations.
  In the real  space these nonstationary solutions for the congruence $K$
  correspond to  a moving
  singular ring, the displacement and orientation of which depend  on the
  real and imaginary parts of the complex world line $X_{o}(\tau _{L})$.
  \bigskip
  {\bf 5.2 Super-extension.}
  \medskip
  \par
  One can extend complex time parameter on the complex superspace by
  introducing the (2,0)- world-sheet supersymmetry. The complex (2,0)-
  supertime coordinates [22] $\tau, \bt  , \xi ^{+},  \bxi ^{+}$
  may be unite in the ciral  superfield,
  $$ {\cT }= \tau  + \eta _{+}(\tau ,\bt )\xi ^{+} + i \xi ^{+}\bxi^{+},
  \eqno(5.8)$$
  and then the analytical complex world-line turns out to be the chiral
  superfunction of supertime,
  $$ {\cX } ^{i} ({\cT }) = x^{i} ( \tau)  + \eta _{+}^{i} (\tau ,\bt
  )\xi ^{+} + i \xi ^{+} \bxi ^{+}
  \partial _\tau x^{i} (\tau ).  \eqno(5.9)$$
  The (2,0)-supercovariant derivatives are $$ D_{+} =
  \partial _{ \xi ^{+}}+i\bxi ^{+}\partial _{\tau} ;\qquad \bD _{+} =
  \partial _{\bxi ^{+} } +
  i\xi ^{+}\partial _{\tau} . \eqno(5.10)$$
  The superfield action corresponding to
  (3.1) takes the form $$ S = {-1\over 4 \pi \alpha '} \int  d\tau
  d\bt d\xi d\bxi  \eta _{i \bar j} D_{+} {\cX } ^{i}_{o}
  \overline{(D_{+} {\cX } _{o}^{j})}
  = $$ \par $$ = {-1\over\pi \alpha '} \int d\tau
  d\bt [(\partial _{\tau }x_{oi})(\overline{\partial_{\tau }x_{o}^{i}})
  -i(\bar {\eta }_{i+}\partial _{\tau} \eta ^{i}_{+}-\overline{ \bar {\eta }
  _{i+} \partial _{\tau }\eta ^{i}_{+}})] \eqno(5.11)$$
  \noindent and gives rise to the Klein-Gordon action for $x^{i}_{o}$  and a
  Dirac action for $\eta ^{i}_{+}.$
  \par
  There  exists a hint that the  problem of the nonstationary
  generalization  of the Kerr - Newman solution and the problem of its
  superextension are  related problems.
  \par
  It is known that the dynamical extension of the Kerr-Newman solution
  runs into many serious problems. One of the obstacles is connected
  with the radiative character of such solutions and in consequence with the
  corresponding violation of the flat asymptotic behavior of the space-time,
  since the stress tensor does not decrease rapidly enough at the large
  distances. One can hope that the corresponding supersolutions could be
  asymptotically flat, because in the vacuum region,
  at large distances from source, the supersymmetry could be restored.
  This ought to guarantee both the vanishing of the stress tensor and the
  asymptotically flat behavior at infinity, in spite of the radiative
  character of nonstationary solutions.
  \par
  In conclusion it is pleasure for me to  thank Professors R.Kerr and
  Z.Perj\`es for  the  kind hospitality  in  the  Central Research
  Institute for Physics of Budapest and for  the very important
  discussions.  I am  thankful also for  the useful conversation to
   V.N. Ponomariov, A.O.Barvinsky, O.A.  Soloviov, D.P. Sorokin.
  \vfill\eject
  {\bf REFERENCES}
  \medskip
  \item{[1]} Witten E.: String theory and black holes. Phys.Rev.{\bf
  D44}, 314 -324 (1991) \medskip
  \item{[2]} Debney G.C., Kerr R.P., Schild A.: J.Math.Phys. Solutions
  of the Einstein and Einstein - Maxwell Equations. {\bf 10}, 1842 -
  1854 (1969) \medskip
  \item{[3]} L\'opez C.A.: Extended model of the electron in general
  relativity. Phys.  Rev.  {\bf D30} 313 - 316 (1984)
  \medskip
  \item{[4]} Israel W.:  Source of Kerr metric.  Phys.  Rev.  {\bf
  D2}, 641 (1970)
  \item{}
  Burinskii A.Ya.: The problem of the source of the Kerr - Newman
  metric:  The volume Casimir effect and superdense pseudovacuum
  state.  Phys.  Lett.{\bf B 216}, 123 - 126 (1989)
  \item{}
  Burinskii A.Ya.: On the problem of the source of the Kerr metric.
  Izvestiya Vuzov Fiz.{\bf 5}, 80 - 86 (1988) (in russian)
  \item{}
  Tiomno I.: Electromagnetic field of rotating charged bodies.  Phys.
  Rev.{\bf D7}, 992 - 997 (1973)
  \item{}
  Hamity V.:  Interior of Kerr metric. Phys. Lett. {\bf A56}, 77 - 78
  (1976)
  \medskip
  \item{[5]} Carter B.: Global structure of the Kerr family of
  gravitational fields.  Phys. Rev. {\bf 174}, 1559 - 1571 (1968)
  \item{}
  Burinskii A.Ya.: Microgeons with spin. Sov. Phys.  JETP {\bf 39}
  193 - 195 (1974)
  \medskip
  \item{[6]} Ivanenko D. and Burinskii A.Ya.: Spin - strings in
  gravity.  Izvestiya Vuzov Fiz.,{\bf 7}, 113 - 119 (1978);
  Gravitational strings in the models of elementary particles.  {\bf 5},
  135 - 137 (1975) (in russian).
  \item{}
  Burinskii A.Ya.: Strings in the Kerr - Schild metrics. In:
  Problem of theory of gravitation and elementary particles {\bf 11} 47
  - 60 (1980), ed.  K.P. Stanyukovich, Moscow, Atomizdat, (in russian).
  \medskip
  \item{[7]} Nielsen H.B. and Olesen P.: Vortex-line models for dual
  strings.  Nucl. Phys.{\bf B61} 45 - 61 (1973)
  \item{}
  Witten E.: Superconducting strings. Nucl.Phys., {\bf B249}, 557 - 592
  (1985)
  \medskip
  \item{[8]} Newman E.T.: Maxwell equations and complex Minkowski
  space. J.Math.  Phys. {\bf 14}, 102 - 103 (1973); Lienard-Wiechert
  fields and general relativity. J.Math.Phys.{\bf 15} 44 - 45 (1974)
  \item{}
  Lind R.W., Newman E.T.: Complexification of the
  algebraically special gravitational fields. J. Math. Phys. {\bf15}
  1103 - 1112 (1974)
  \medskip
  \item{[9]} Belavin A.A., Polyakov A.M. and Zamolodchikov A.B.:
  Infinite conformal symmetry in two-dimensional quantum field theory.
  Nucl. Phys. {\bf B241}, 333 - 380 (1984)
  \medskip
  \item{[10]} Shaw W.T.: An ambitwistor description of bosonic or
  supersymmetric minimal surfaces and strings. Class.Quant.Grav.
  {\bf3}, 753 - 761 (1986)
  \item{}
  Burinskii A.Ya.: Kerr's Particle With Spin and Superstrings.
  In: Abstracts of the International Symposium  on  Supernovae  and
  High Energy Astrophysics, December 27, p.10, 1989, Calcutta, India.
  \medskip
  \item{[11]} Hamidi S., Vafa C.: Interactions on orbifolds. Nucl.
  Phys.  {\bf B279}, 465 - 513 (1987) \medskip
  \item{[12]} Ooguri H., C. Vafa C.: Geometry of N = 2 strings. Nucl.
  Phys.{\bf B 361}, 469 - 518 (1991); N = 2 heterotic strings.{\bf B
  367} 83 - 104 (1991) \medskip
  \item{[13]}  Veblen O.: Geometry of two-component spinors.
  Proc.Nat.Acad.Sci.  USA,{\bf XIX}, 462 (1933)
  \item{}
  Ruse H.: On the geometry of Dirac's equations and their expression in
  tensor form. Proc.Roy.Soc. of Edinburg {\bf 37}, 97 - 127 (1936/37)
  \medskip
  \item{[14]} Dixon L., Harvey J.A., Vafa C., Witten E.: Strings on
  orbifolds.  Nucl. Phys. {\bf B261}, 678 -686 (1985); String on
  orbifolds (II). {\bf B274} 285 -314 (1986) \medskip
  \item{[15]} Green M.B., Schwarz J.H., Witten E.: Superstring
  Theory,v.1,2, Cambrige Univ. Press., Cambrige, 1988.  \medskip
  \item{[16]} Latorre J.I., L\"utken C.A.: Constrained $ CP^n $ model.
  Phys. Lett. {\bf B 222}, 55 - 60 (1989)
  \item{}
  Gates S.J.,Jr., H\"ubsch T.: Calabi-Yau heterotic strings and
  unidexterous $\sigma $ - models. Nucl.Phys. {\bf B 343}, 741 - 774
  (1990)
  \item{}
  Greene B.R., Vafa C. and Warner N.P.: Calabi - Yau
  manifolds and renormalization group flows. Nucl.Phys. {\bf B 324},
  371 - 390 (1989)
  \item{}
  Martinec E.J.: Algebraic geometry and effective
  Lagrangians. Phys.  Lett. {\bf B 217}, 431 - 436 (1989)
  \medskip
  \item{[17]} Candelas P., Green P.S., H\"ubsch T.: Rolling among
  Calabi - Yau vacua. Nucl.Phys.  {\bf B330}, 49 - 102 (1990) \medskip
  \item{[18]} Gepner D.:
  Exactly solvable string compactifications on manifolds of SU(N)
  holonomy.  Phys. Lett.  {\bf B 199}, 380 - 387 (1987)
  \medskip
  \item{[19]} Penrose R.:  Twistor algebra.  J. Math. Phys.{\bf 8},
  345 - 366 (1967) \medskip
  \item{[20]} Burinskii A., Kerr R.P. and Perj\`es Z.:
  Nonstationary generalization of the Kerr congruence (in preparation),
  \medskip
  \item{[21]} Kerr R.P., Willson W.B.:  Singularities in the Kerr -
  Schild metrics. Gen.  Rel.  Grav.:  {\bf 10}, 273 - 281 (1979)
  \medskip
  \item{[22]} Brooks R., Muhammad F., Gates S.J., Jr.:  Unidexterous
  D=2 supersymmetry in superspace.  Nucl.  Phys.  {\bf B268}, 599 - 620
  (1986)
  \item{}
  Hull C.M., Witten E.:  Supersymmetric sigma models and
  the heterotic string. Phys. Lett.  {\bf B 160}, 398 - 402 (1985)
  \item{}
  Hull C. M., Spence B.: The (2,0) supersymmetric Wess - Zumino -
  Witten model.  Nucl. Phys. {\bf B 345}, 493 - 508 (1990)
  \enddocument
  \end

  \end